\documentclass[a4paper,12pt]{article}
\title{General existence proof for rest frame systems in asymptotically flat space-time}
\author{Sergio Dain\thanks{Fellowship holder  of CONICOR} \thanks{E-mail: dain@fis.uncor.edu} \thanks{Present address: Max-Planck-Institut f\"ur Gravitationsphysik, Am M\"uhlenberg 1, D-14476 Golm, Germany} \, and Osvaldo M. Moreschi\thanks{Member of CONICET}\\
FaMAF, Ciudad Universitaria, Universidad Nacional de C\'ordoba, \\ (5000) C\'ordoba, Argentina}
\usepackage{amsmath,amssymb,amsfonts}
\newtheorem{theorem}{Theorem} [section]
\newtheorem{lemma}{Lemma}[section]

\begin{document}
\maketitle
\begin{abstract}
We report  a new result on the nice section construction used in the definition of rest frame
systems in general relativity. This construction is needed in the study of non trivial 
gravitational radiating systems.   
We prove existence, regularity and non-self-crossing property of solutions of the 
nice section  equation 
for general asymptotically flat space times. This proves
 a conjecture enunciated in a previous work.

\vspace{0.5cm}

\end{abstract}

\section{Introduction}

The rest frame equation, or ``nice section'' equation\cite{Moreschi88}, for asymptotically flat 
space-times, is an equation which relates  quantities defined at null infinity (scri or $\mathcal{I}$).
 The solutions of this equation are a family of privileged sections of scri, which represent the 
analog of having a frame of reference\cite{Moreschi98}\cite{Erratum99}. 

The physical motivations for  the nice section construction have been explained already
in the literature\cite{Moreschi88}\cite{Moreschi98}. We just mention here that at present the 
description of astrophysical gravitational radiation in terms of the sources, is carried out through  the
so called quadrupole radiation formula\cite{Landau59}. The construction of the nice sections 
provides a necessary tool, that we expect to be useful for the discussion of gravitational 
radiation and its relation with the notion of multipole moments, including the cases
of systems which suffer a non-trivial back reaction due to gravitational radiation, or when
the lapse of time considered is relatively long, when compared to the dynamical characteristic time. 
Those cases can not be dealt with the quadrupole radiation formula, which has been deduced
for  an instant of time, and for weak fields. 

In what follows we will concentrate in the mathematical
aspects of the nice section construction, and we will provide a general existence proof of them;
which demonstrates that the conjecture that we have enunciated in ref. \cite{Moreschi98} is true.

Let us start by recalling the basic structure of scri. Roughly speaking scri is the three 
dimensional manifold that corresponds to the end points when one takes the limits to infinity of  the future directed null geodesic. 
The topology of a complete scri is $\mathbb{R} \times S^2$, and the natural coordinates 
 are the so called Bondi coordinates; which are labeled by    $(u,\zeta, \bar \zeta)$, 
where $(\zeta, \bar \zeta)$ are complex stereographic coordinates of the sphere $S^2$, 
and $u$ takes  values in $\mathbb{R}$.
Given   an arbitrary real function $\gamma(\zeta, \bar \zeta)$ of the sphere, 
we say that $u=\gamma(\zeta, \bar \zeta)$ defines  a section of $\mathcal{I}$. 

The symmetry group of scri (i.e. the group of coordinate transformations which preserves the asymptotic geometric  structure of the space-time) is the  Bondi-Metzner-Sach (BMS) group. It  is defined by the following relations:
\begin{equation} \label{BMS}
\tilde u =K(u-\gamma(\zeta,\bar\zeta)), \quad \tilde \zeta=
\frac{a\zeta +b}{c\bar \zeta+d},
\end{equation}
with
\[
ad-bc=1, \quad K=\frac{1+\zeta \bar \zeta}{|a\zeta+b|^2+|c\zeta+d|^2}
\]
where $a,b,c,d$ are complex constant, and $\gamma$ is a  real  function on the sphere.

We express the supermomentum at a section $u=\text{constant}$ in the following way:
\begin{equation} \label{supermomentum}
P_{lm}=-\frac{1}{\sqrt{4\pi}} \int Y_{lm}(\zeta, \bar \zeta) \Psi(u,\zeta,\bar \zeta) dS^2, 
\end{equation}
where $dS^2$ is the surface element of the unit sphere; $Y_{lm}$ are the spherical harmonics;
\begin{equation} \label{Psi}
\Psi=\Psi_2 + \sigma \dot{\bar \sigma} +\eth^2 \bar \sigma;
\end{equation}
with $\Psi_2 $   and    $\sigma$
 being the leading order asymptotic behavior of the second 
Weyl tensor component and the Bondi shear respectively;  where 
we are using the GHP notation\cite{Geroch73} for the edth operator of the 
unit sphere, and an   over dot means $\partial /\partial u$.

Similarly the total Bondi energy-momentum vector at a given section of null infinity can be expressed by
\begin{equation} \label{P}
P^a=-\frac{1}{4\pi} \int l^a(\zeta, \bar \zeta) \Psi(u,\zeta,\bar \zeta) dS^2, 
\end{equation}
where
\begin{equation} \label{la}
 (l^a) =\left( 1,\frac{\zeta +\bar \zeta }{
1+\zeta \bar{\zeta }},\frac{\zeta -\bar{\zeta }}{i(1+\zeta 
\bar{\zeta )}},\frac{\zeta \bar{\zeta }-1}{1+\zeta \bar{\zeta 
}}\right).
\end{equation}
In fact the Bondi momentum is a linear combination of the first four component $(l=0,1)$  of the 
supermomentum.

The rest mass $M$ of the section  is given by
\begin{equation} \label{M}
M=\sqrt{P^a P_a},
\end{equation}
where the indices are moved with the Lorenzian flat metric $\eta_{ab}$ at scri\cite{Moreschi86}.
 
The BMS group (\ref{BMS}) and the supermomentum  (\ref{supermomentum}) are the basic 
ingredients for the definition of the nice section equation. 

A rest frame system in special relativity  is a coordinate system in which the total four momentum has only temporal component. Given such a system we can perform a translation and we still have a rest frame system.

This idea can be translated to the structure of scri; where the supermomentum takes the role of the 
momentum in special relativity. With this idea in mind, the nice sections are defined in such a way 
that the supermomentum has only temporal component. This means that  the integrand $\Psi$ at a 
nice section is a constant, i.e., it does not depend on the angular coordinates $(\zeta, \bar \zeta)$.

The function $\Psi$ transforms under a BMS transformation in the following way\cite{Moreschi88}  
\begin{equation} \label{BMSPsi}
\tilde \Psi= \frac{1}{K^3}(\Psi-\eth^2 \bar \eth^2 \gamma),
\end{equation}
where $\tilde \Psi =\tilde \Psi_2 + \tilde \sigma \dot{\bar{\tilde \sigma}}+{\tilde {\eth}}^2 \bar{\tilde \sigma}$.

This means that  given an initial arbitrary  section, a nice section can be determined
 by the appropriate 
BMS transformation such that $\tilde \Psi$ is constant. Therefore we can express the nice section equation by 
\begin{equation} \label{nice}
\eth^2 \bar \eth^2  \gamma =\Psi(u=\gamma,\zeta,\bar \zeta) +K^3(\gamma,\zeta, \bar \zeta) M(\gamma),
\end{equation}
where the conformal factor   $K$ can be related to the Bondi momentum by 
\begin{equation} \label{K}
K=\frac{M}{P^a l_a},
\end{equation}
and $P^a$ is evaluated at  the section $u=\gamma$; which is calculated through the integral (\ref{P}).
 
Since the operator $\eth^2 \bar \eth^2$ occurs very frequently in this work we use the notation $D=\eth^2 \bar \eth^2$, and we note also that 
\begin{equation} \label{D}
D\equiv \eth^{2} \bar \eth^{2} =\frac{1}{4} \Delta ^{2} -\frac{1}{2} \Delta
\end{equation}
where  $\Delta $ is the Laplacian of the unit sphere.

We say that a supertranslation $\gamma$ (or equivalently a section of $\mathcal{I}$) satisfies the ``nice section'' equation if the function $\tilde \Psi$ evaluated in $\gamma$ is equal to  the total mass $M$; this is exactly the content of equation (\ref{nice}). 

Note that the  nice section equation involves also a Lorenz rotation, as it can  be seen by 
the appearance of the scalar  $K$ in the equation.

Equation (\ref{nice}) is our fundamental equation, we want to investigate the existence and properties of solutions of this equation. 

We say that a real function $x=x(\zeta, \bar \zeta)$ on the sphere is a translation if $Dx=0$. Note that this implies that $x$ has an expansion in spherical harmonics with $l=0,1$. An  arbitrary 
 regular function $\gamma$ can be decomposed in 
\[
\gamma=x+y,
\]
where $y$ has an expansion with $l\geq 2$. Given an arbitrary $x$, equation (\ref{nice}) is an equation for $y$.

The main result of this paper is the following Theorem:

\begin{theorem} \label{maintheorem}
If  $\Psi$ is a smooth function on scri,  the total energy $P^0$ is bounded by the constant $E_0$, 
the total mass $M$ is bounded from below by $M\geq M_0$, $M_0 >0$; and the gravitational energy density flux $|\dot \sigma|^2 \leq \lambda$; where the constant $\lambda$ satisfies
\[
\lambda <\frac{\sqrt{27}}{4}(1+(2C_K)^4)^{-1},
\] 
and $C_K$ is given by
\[
C_K=\frac{E_0}{M_0}+\sqrt{\frac{E^2_0}{M^2_0}-1};
\]
then

(i) For every translation $x$ there exists a solution $y$ of  equation (\ref{nice}), and $y$ is a smooth function on the sphere.

(ii) The solutions $\gamma=x+y(x)$ are continuous in the 4-parameter translation $x$, and if $x_1$ and $x_2$ are two translations such that the difference, $x_2-x_1$,  corresponds to  a future directed time like vector, then
\[
\gamma(x_1,\zeta, \bar \zeta)< \gamma(x_2,\zeta, \bar \zeta).
\]
\end{theorem}
Since the mass $M$ is a decreasing function of $u$, the constant $M_0$ is the final rest mass as $u\rightarrow \infty$.
Note that while $M_0$ is Lorentz invariant $P^0$ is not; in other words the inequalities depend on the particular Bondi system being used.

It is important to recall that all the hypothesis of the theorem are in terms of physical quantities; namely the total mass $M$, the energy $P^0$ and the gravitational energy density flux $|\dot\sigma|^2 $. Theorem \ref{maintheorem} essentially says that the  solutions of equation (\ref{nice}) exist and have the expected physical properties, when  the gravitational radiation of the space-time is not too high.  

%%% New Paragraph
As it was pointed out in \cite{Moreschi98}, there exist  another
choices for the supermomentum $\Psi$ in the literature. We want to
analyze here these other possibilities.  
One of them is  the Geroch supermomentum\cite{Geroch77}\cite{Dray84}
\[
\Psi_{G}=\Psi_2 + \sigma \dot{\bar \sigma} +\frac{1}{2}(\eth^2
  \bar \sigma-\bar \eth^2  \sigma);
\]
and another one is  the Geroch-Winicour supermomentum\cite{Geroch81}\footnote{We take this opportunity to
  correct a mistake in the  articles \cite{Walker83} and \cite{Moreschi88}, the
  supermomentum $\Psi$ \emph{is not} the same as the Geroch-Winicour
  supermomentum, as it is clear from the expression (\ref{eq:GW})} 
\begin{equation}
  \label{eq:GW}
 \Psi_{GW}=\Psi_2 + \sigma \dot{\bar \sigma} -\bar \eth^2  \sigma.  
\end{equation}

The function $\Psi_G$ is invariant under supertranslation, then it
will not give an equation for $\gamma$. Note that  $\Psi_G$ vanishes 
for Minkowski space-time, but equation (\ref{nice}) is not trivial
in this space-time; its solutions are the shear-free sections ($\sigma=0$).
As it is well known, the sections  with $\sigma=0$ in
Minkowski space-time represent  a natural way 
to isolate the Poincare group from the BMS group of scri, and this is 
 precisely the aim of the `nice section' equation in the general
 case. These considerations rule out $\Psi_G$.

The Geroch-Winicour supermomentum  $\Psi_{GW}$ could in principle be used to
construct an alternative equation. But the supermomentum $\Psi$
satisfies the following  remarkable equation 
\begin{equation} \label{Psidot}
\dot \Psi=|\dot \sigma|^2,
\end{equation}
which relates in a simple way the time derivative of $\Psi$ with the
gravitational radiation flux. 
This equation  allows us to write the hypothesis of the theorems in
terms only of  the total mass $M$, the energy $P^0$ and the
gravitational energy density flux $|\dot\sigma|^2 $.  It might be also possible to prove
similar theorems for an equation based on $\Psi_{GW}$, but they  would
certainly look much more complicated.

%%%%%%%%%%%

In section \ref{existencesection} we prove the existence part (i) of the theorem. In section \ref{uniquenesssection} we prove the part (ii). Finally, for the sake of completeness, we remind in the Appendix the Schauder fixed  point theorem, which is the main tool for the existence proof of section \ref{existencesection}.

\section{Existence Proof} \label{existencesection}

The equation (\ref{nice}) is an  elliptic, nonlinear, equation on the compact manifold $S^2$. The main tool for proving existence of solutions of this type of equations is the Schauder fixed point Theorem \ref{Schauder} (see for example  page 380 of ref. \cite{McOwen96} for an elementary treatment of  a similar equation with the same technique). 

Let us define the function $f$ as the right-hand side of equation (\ref{nice})
\begin{equation} \label{f}
f(x+y,\zeta,\bar \zeta)=\Psi(x+y,\zeta,\bar \zeta)+K^3  M(x+y,\zeta,\bar \zeta).
\end{equation}
The first step  is to prove that the function $f$ satisfies the
appropriate inequality under the hypothesis of Theorem
\ref{maintheorem}; this will be proved in the following Lemma.

%%%% I have cut the equation Psidot

\begin{lemma} \label{boundf}
Assume that $\Psi$ is a smooth function on scri, the gravitational radiation flux satisfies $|\dot \sigma|^2 \leq \lambda$ ,  the total mass $ M\geq M_0>0$ and the total energy $P^0 \leq E_0$. Then 
\begin{equation} \label{eboundf}
|f| \leq \lambda |y| +C,
\end{equation}
where the constant $C$ does  not depend on $y$.
\end{lemma}

\emph{ Proof:}
Take $u_0$ an arbitrary, fixed, real number.  From equation (\ref{Psidot}) we have
\[
\Psi (\gamma, \zeta, \bar \zeta)= \int_{u_0}^\gamma  |\dot \sigma|^2 (u', \zeta, \bar \zeta) du'+ \Psi (u_0, \zeta,\bar \zeta).
\]
We use the hypothesis  $|\dot \sigma|^2 \leq \lambda$ to obtain 
\begin{equation} \label{boundPsi}
|\Psi| \leq \lambda |y| +C_1,
\end{equation}
where 
\[
C_1= \sup_{S^2} \{ \lambda|x- u_0|+ |\Psi (u_0, \zeta, \bar \zeta)| \}.
\]
Note that the constant $C_1$ depends on $x$ and $u_0$ but not on $y$.

Looking at  the expression (\ref{f}) we see that for proving the inequality (\ref{eboundf}) we need
 only to find  bounds for $K$ and $M$ in terms of the constant $M_0$ and $E_0$. The last one follows immediately since $M \leq P^0$ and by hypothesis we have $P^0\leq E_0$, then
\begin{equation}\label{boundM}
M\leq E_0.
\end{equation} 

Define the velocity vector $V^a$ by $V^a = P^a/M$. Note that  
$V^aV_a=1$.
We use the definition of $K$, equation (\ref{K}), to obtain 
\begin{equation} \label{KV}
K=\frac{1}{V^a l_a}.
\end{equation}
The function $ V^a l_a$ satisfies the following inequality
\[
|V^a l_a| \geq |V^0| -|V^i l_i| \geq |V^0| -\sqrt{V^i V_i}.
\]
We have used that $l^i l_i =1$, with $i=1,2,3$. Then we use that $(V^0)^2 - V^i V_i=1$ to obtain
\begin{equation}\label{Vl}
|V^a l_a| \geq |V^0|-\sqrt{(V^0)^2-1}.
\end{equation}
The right-hand side of this inequality is a  monotonically decreasing function of $|V^0|$. By hypothesis    we have  the bound 
\begin{equation} \label{V0}
V^0=\frac{P^0}{M} \leq \frac{E_0}{M_0}.
\end{equation}
We use the bound (\ref{V0}) in equation (\ref{Vl}) 
to obtain 
\[
|V^a l_a| \geq \frac{C_0}{M_0}-\sqrt{\frac{C_0^2}{M^2_0}-1} ;
\]
and then by equation (\ref{KV}) we get the desired bound for $K$:
\begin{equation} \label{boundK}
|K| \leq C_K,
\end{equation}
where $C_K$ is given by 
\[
C_K=\left( \frac {E_0}{M_0}-\sqrt{\frac{E_0^2}{M^2_0}-1} \right) ^{-1}=\frac {E_0}{M_0}+\sqrt{\frac{E_0^2}{M^2_0}-1}.
\]
Using the bounds (\ref{boundPsi}), (\ref{boundM}) and (\ref{boundK}) we obtain the final inequality for the function $f$:
\[
|f| \leq \lambda|y|+C,
\]
where
\[
C=C_1+E_0 C_K^3.
\] 
$\blacksquare$

With this Lemma we are in position to prove the following existence Theorem: 

\begin{theorem} \label{Existence}
Assume that $\Psi$ is a smooth function on scri,  the total energy $P^0\leq E_0$, the total mass $M\geq M_0>0$ and the gravitational energy density flux $ |\dot \sigma|^2 \leq \lambda$ with
\[
\lambda < \sqrt{\frac{27}{4}}. 
\]
Then for every translation $x$ there exists  a solution $y$ of  equation (\ref{nice}), and $y$ is a smooth function on the sphere. 
\end{theorem}

\emph{ Proof:} Consider the mapping $A$  on the space of continuous functions on the sphere $A: C^0(S^2) \rightarrow C^0(S^2)$ defined as follows
\[
A(y)=D^{-1} f(x+y, \zeta, \bar \zeta),
\]
where $D^{-1}$ is the inverse of $D$. First note that $A$ is well defined for  every translation $x$, since for an arbitrary $x+y$ the function $f$  is orthogonal to the kernel of $D$, and then it  is in the domain of $D^{-1}$. This is not obvious and it is essential, otherwise   equation  (\ref{nice}) would be inconsistent. For a proof see \cite{Moreschi98}.

A solution of  equation (\ref{nice}) is a fixed point $y=A(y)$ of the mapping $A$. To prove that such fixed point exists  we will use the Schauder fixed point Theorem \ref{Schauder}. In the following we will prove that the mapping $A$ satisfies the hypothesis of this Theorem.

First we prove that $A$ is continuous in the $C^0$ norm. By definition 
\[
DA(y)=f(x+y, \zeta, \bar \zeta),
\]
we take  the $L^2$ norm in both sides of this equation
\[
||DA(y)||_{L^2}=||f(x+y, \zeta, \bar \zeta)||_{L^2}\leq \sqrt{4\pi} ||f(x+y, \zeta, \bar \zeta)||_{C^0}.
\]
The last inequality follows because $f$ is continuous. We use the inequality\cite{Moreschi98}
\begin{equation} \label{iD}
 ||Dy||_{L^2} \geq \sqrt{4\pi} \sqrt{27/4} ||y||_{C^0},
\end{equation}
which is nothing but a particular case of the elliptic regularity inequalities for the elliptic operator $D$, to obtain
\begin{equation}\label{contA}
||A(y)||_{C^0} \leq  \sqrt{4/27} ||f(x+y, \zeta, \bar \zeta)||_{C^0}.
\end{equation}
Since $D^{-1}$ is a linear operator we obtain also  the inequality
\begin{equation} \label{Acontinuous}
||A(y_1)-A(y_2)||_{C^0}=||D^{-1} (f(y_1)-f(y_2))||_{C^0} \leq \sqrt{\frac{4}{27}} ||(f(y_1)-f(y_2))||_{C^0}.
\end{equation}
Since by hypothesis $f$ is continuous as a function of  $y$ then by (\ref{Acontinuous}) $A$ is also continuous.

Let $B$ be the closed ball in $C^0(S^2)$ of radius $C_2$; $B$  is clearly a closed, convex subset of the Banach space $C^0(S^2)$. We want to prove that $A$ maps $B$ into itself when  the constant $C_2$ is
 appropriately  chosen. Take $y \in B$, using the inequality (\ref{contA}) and Lemma \ref{boundf} we have
\begin{equation} \label{inqA}
||A(y)||_{C^0}\leq \sqrt{\frac{4}{27}}(\lambda |y| +C)\leq \sqrt{\frac{4}{27}}(\lambda C_2 +C).
\end{equation}
Take $C_2 = \alpha C$, where $\alpha$ satisfies 
\[
\lambda + \frac{1}{\alpha}\leq \sqrt{\frac{27}{4}};
\]
this $\alpha$ exists because we have assumed that $\lambda <\sqrt{\frac{27}{4}}$. Then from the inequality (\ref{inqA}) we obtain
\[
||A(y)||_{C^0}\leq C_2.
\]  
Therefore we have proved that $A$ maps the ball $B$ into itself.

It remains to be proved that the set $A(B)$ is precompact. The set $A(B)$ is a subset of  $H^4(S^2)$. This is a consequence of the elliptic character of the  4th order operator $D$ and the standard elliptic regularity theorems (see for example \cite{Aubin82} \cite{Morrey66})

Since $H^4(S^2)$ is compactly imbedded in $C^0(S^2)$ (see for example \cite{Gilbarg}) it follows  that $A(B)$ is precompact in $C^0(S^2)$. 

The hypothesis of Theorem \ref{Schauder} are satisfied, and then there exists  a solution $y=y(\zeta,\bar \zeta)$ in $C^0(S^2)$.

Since $y$ is in $C^0(S^2)$ and $f$ is smooth, then $f(x+y, \zeta, \bar \zeta)$ is in $L^2(S^2)$. 
Therefore by the elliptic regularity theorems  $y \in H^4 (S^2)$; and then, by the Sobolev imbedding Theorem, it is in $C^2(S^2)$;  but then $f(x+y, \zeta, \bar \zeta)$ is in $H^2(S^2)$. We use 
elliptic regularity and induction to conclude that  $y$ is in $C^\infty(S^2)$. $\blacksquare$

Note that for the existence proof we only need $\Psi$ to be in $C^0$; if we only require this, then   by the elliptic regularity  theory 
 one would only obtain a $C^2$ solution $y$.

\section{Uniqueness and non self-crossing properties of the solutions} \label{uniquenesssection}

Here we want to analyze the dependence of the solution $y(x,\zeta,\bar \zeta)$ in terms of the translation $x$.  In the previous section we have shown  that given an arbitrary  $x$ there exists  a solution $y(x,\zeta,\bar \zeta)$, which is smooth in the  variables $(\zeta,\bar \zeta)$. Now we want to prove that under more restrictive conditions this solution is unique. Moreover, we also want to prove that $y(x,\zeta,\bar \zeta)$ is continuous in $x$. This  means that  the function $y(x,\zeta,\bar \zeta)$ 
defines  a family of solutions parameterized by $x$. We also prove that this family has the following  non self-crossing property: if $x_1$ and $x_2$ are two translations such that the difference $x_2-x_1$ corresponds to  a future directed timelike vector,  then 
\[
\gamma(x_1,\zeta,\bar \zeta)<\gamma(x_2,\zeta,\bar \zeta).
\]
This implies that the solutions of  equation (\ref{nice}) generated by a time translation do not cross among them. 

We begin with the following auxiliary Lemma.

\begin{lemma} \label{boundf'} 
Assume that the hypothesis of Lemma \ref{boundf}  holds.  Then
\[
||f(\gamma_2)-f(\gamma_1)||_{L^2} \leq \sqrt{4\pi} C_{f'}||\gamma_2- \gamma_1||_{C^0},
\]
where 
\begin{equation} \label{Cf'}
C_{f'}=\lambda \left( 1+(2C_K)^4 \right) .
\end{equation}
\end{lemma}
\emph{ Proof:}
Consider $f$ as a mapping $f: C^0(S^2)\rightarrow L^2(S^2)$,  since $\Psi$ is smooth then $f$ is Fr\'echet differentiable, and by the mean value theorem (see for example \cite{Choquet77}
) we obtain the following bound
\[
||f(\gamma_2)-f(\gamma_1)||_{L^2} \leq \sup_{t\in (0,1)}||f'(\gamma_1 +t(\gamma_2-\gamma_1)||_{{\cal L} (C^0,L^2)} ||\gamma_2-\gamma_1||_{C^0},
\]
where ${{\cal L} (C^0,L^2)}$ denotes the operator norm, and $f'$ is the Fr\'echet derivative of $f$.
 We have to calculate the bound for $f'$. The  Fr\'echet derivative $f'$  is given 
explicitly by\cite{Moreschi98} 
\[
f'(\gamma) \delta \gamma = \dot \Psi(\gamma) \delta \gamma  
+\left(\frac{4P^a}{M}-3Kl^a \right) K^3 \delta P_a,
\]
where
\[
\delta P_a=-\frac{1}{4\pi} \int l_a \dot \Psi(\gamma) \delta \gamma dS .
\]
And the operator norm of $f'$ is 
\[
||f'(\gamma)||_{{\cal L} (C^0,L^2)}=\sup \{ ||f'(\gamma)\delta \gamma ||_{L^2};  ||\delta \gamma ||_{C^0}=1 \}.
\]

We use  the bounds $E_0, M_0, \lambda$ and the bound (\ref{boundK}) for $K$ to obtain
\[
 ||f'(\gamma)||_{{\cal L} (C^0,L^2)}\leq \sqrt{4\pi} C_{f'},
\]
where
\[
C_{f'}=\lambda \left( 1+(2C_K)^4 \right),
\]
and we have used the inequality
\[
\frac{1}{C_K}\leq K \leq C_K.
\]
$\blacksquare$

\begin{theorem} \label{Family}
Assume that the hypotheses of Theorem \ref{Existence} hold, and  in addition that
\begin{equation} \label{lambdainq}
\lambda<\frac{\sqrt{27}}{4}(1+(2C_K)^4)^{-1}. 
\end{equation}
Then the solutions $\gamma =x+y(x)$ are continuous in the 
4-parameter translation $x$. Moreover, if $x_1$ and $x_2$ are two translations such the difference $x_2-x_1$ corresponds to  a future directed time like vector,
equivalently $x_2-x_1>0$;  then 
\[
\gamma(x_1,\zeta,\bar \zeta ) < \gamma(x_2,\zeta , \bar \zeta),
\]
for all $(\zeta , \bar \zeta)$ in the sphere.
\end{theorem}

\emph{ Proof:}
Let  $x_1+y(x_1,\zeta,\bar \zeta)$ and $x_2+y(x_2,\zeta,\bar \zeta)$ be two solutions  of the equation (\ref{nice}); taking the difference we obtain
\[
D(y(x_2)-y(x_1))=f(x_2+y(x_2),\zeta,\bar \zeta)-f(x_1+y(x_1),\zeta,\bar \zeta).
\]
Let us  take the $L^2$ norm on  both sides of this equation
\[
||D(y(x_2)-y(x_1))||_{L^2}=||f(x_2+y(x_2),\zeta,\bar \zeta)-f(x_1+y(x_1),\zeta,\bar \zeta)||_{L^2},
\]
then  using Lemma \ref{boundf'} and the inequality (\ref{iD})  we  obtain
\[
\sqrt{27/4}||y(x_2)-y(x_1)||_{C^0}\leq C_{f'}||x_2+y(x_2)-(x_1+y(x_1))||_{C^0}.
\]
Since  by hypothesis we have  inequality (\ref{lambdainq}), it is deduced that   
 $C_{f'} < \sqrt{27}/4$ and we obtain
\[
||y(x_2)-y(x_1)||_{C^0}\leq \frac{C_{f'}}{\sqrt{27/4}-C_{f'}} ||x_2-x_1||_{C^0};
\]
in this way we have proved that the function $y(x)$ is continuous in $x$. Note that 
 \[
\frac{C_{f'}}{\sqrt{27/4}-C_{f'}}<1,
\]
then  for $x_1 < x_2$ we obtain
\begin{equation} \label{timelike}
||y(x_2)-y(x_1)||_{C^0} <  ||x_2-x_1||_{C^0}.
\end{equation}
The fact that  $x_2-x_1$ corresponds to a future directed  timelike vector, is equivalent to the statement that we can choose  a coordinates system (by means of a  Lorentz rotation) in which $x_2-x_1$ is a positive constant. Then $||x_2-x_1||_{C^0}=x_2-x_1$ in this coordinate system and from the inequality (\ref{timelike}) we conclude that
\[
\gamma (x_2,\zeta, \bar \zeta)>\gamma(x_1,\zeta, \bar \zeta).
\]
It is important to remark that after one has proved  the last inequality, it is possible to come 
back to the original Bondi system by the inverse of the original transformation which does not 
change the inequality. That is, this inequality is also true in the original Bondi system.
$\blacksquare$

\section*{Appendix: Schauder Fixed Point Theorem}
We use the following version of the Schauder Theorem (see ref. \cite{Gilbarg} page 280.)
\begin{theorem} \label{Schauder}
Let $B$ be a closed convex set in a Banach space $V$ and let $A$ be a continuous mapping of $B$ into itself such that the image $A(B)$ is precompact (i.e. the closure of $A(B)$ is compact). Then $A$ has a fixed point.
\end{theorem}

\section*{Acknowledgments}
We are deeply indebted  to G. Nagy and O. Reula for  very illuminating discussion about the Schauder fixed  point theorem. 

We acknowledge support from  Fundaci\'on Antorchas,  SeCyT-UNC, CONICET and
FONCYT BID 802/OC-AR PICT: 00223.

\end{document}